# Observation of large positive magnetoresistance and its sign reversal in GdRhGe

Sachin Gupta[a], K.G. Suresh[a,*] and A.K. Nigam[b]

[a]Department of Physics, Indian Institute of Technology Bombay, Mumbai-400076, India

[b]Tata Institute of Fundamental Research, Homi Bhabha Road, Mumbai-400005, India

## Abstract

Magnetic properties, heat capacity and magnetoresistance (MR) of polycrystalline GdRhGe are investigated. It shows two antiferromagnetic transitions, one at $T_1$=31.8 K and the other at $T_2$=24 K, and field induced metamagnetic transition over a wide temperature range. The ac susceptibility data reveal that the transition at 24 K is not simple antiferromagnetic. Dominant contributions to the heat capacity and the resistivity have been identified. MR is found to show sign reversal just below $T_1$ and attains a large positive value of 48% at 2 K for 50 kOe. Like MR, the isothermal magnetic entropy change also undergoes a sign reversal as the temperature is varied, indicating a change of the magnetic structure and the moment amplitude in determining these properties.

*Corresponding author email: suresh@phy.iitb.ac.in

## 1. Introduction

There is considerable interest in the study of artificial multilayer materials composed of the magnetic and nonmagnetic metals. A number of multilayer systems [1-3] has been found to show large decrease in electrical resistivity on the application of a magnetic field and therefore promising for applications. Some bulk intermetallic compounds [4-7] have also been found to show large magnetoresistance (MR). The mechanism of giant magnetoresistance (GMR) in multilayer systems differs from that in bulk intermetallic compounds in which the MR is due to the direct effect of field on the conduction electrons or on scattering impurities. On the other hand, in multilayer systems it arises from the reorientation of single domain magnetic layers [8]. It has been observed that in bulk magnetic materials, GMR generally arises at a field induced magnetic transition, usually associated with a crystallographic transformation or volume change [7]. For example, GMR in FeRh alloy [4] is due to the antiferromagnetic phase made up of the alternating monolayers of Fe (magnetic) and Rh (nonmagnetic) and occurs when the material changes state from antiferromagnetic (AFM) to ferromagnetic (FM) on the application of field. On the other hand, the occurrence of GMR in $Gd_5Si_{1.8}Ge_{2.2}$ is attributed to the variation in the density of states at the Fermi level as a consequence of a crystal symmetry change [6].

The rare earth (R) - transition metal (T) intermetallic compounds, where 3d/4d electrons of transition metals show itinerant behavior and 4f electrons of rare earths are localized, have attracted much interest due to their anomalous magnetic and electrical properties. It is well known that a number of mechanisms are possible in transition metals because the same carriers are responsible for both magnetism and transport phenomena, unlike the rare earths where the magnetism originates mainly from the 4f electrons and the transport is associated with 6s and 5d electrons. Though negative GMR is usual in many such compounds, positive GMR has also been observed in some systems such as $R_2Ni_3Si_5$ (R=Tb,Sm,Nd) [9], $LaMn_2Ge_2$ [10], $GdCu_6$ [11], and pure Nd metal [12].

As part of our search for novel materials with interesting magnetic and related properties, we have studied a series of compounds of the RTX family, where T is a magnetic/nonmagnetic element and X is a p-block element. One of such systems is the RRhGe series, where Rh is a 4d transition metal. In this paper, we report the results of the work carried out on the magnetic and the magneto-transport properties of GdRhGe. Since the magnetocaloric effect (MCE) also helps in understanding the correlation between the magnetic and transport properties in certain materials, we have also estimated MCE as a function of temperature in different fields.

## 2. Experimental details

The polycrystalline GdRhGe compound was synthesized by arc melting the constituent elements with purity better than 99.9% in argon atmosphere. The melted ingot was sealed in evacuated quartz tube and annealed for 8 days at 800℃ for better homogeneity. The room temperature powder x-ray diffraction (XRD) pattern was taken on the X'PERT PRO diffractometer using Cu Kα radiation. DC magnetization and heat capacity C(T) measurements were performed using a Physical Property Measurement System (Quantum Design, PPMS-6500). AC Susceptibility (ACS) measurement was performed on Quantum Design MPMS squid VSM. The electrical resistivity (ρ) measurement with and without field was also performed on PPMS using standard four probe technique, applying an excitation current of 150 mA.

## 3. Results and discussion

The Rietveld refinement of the XRD pattern collected at room temperature shows that the compound is single phase and crystallizes in TiNiSi type orthorhombic structure with space group Pnma (No. 62). It is reported that TiNiSi-type structure is the superstructure of $Co_2Si$-type structure [13]. The characteristic feature of this lattice is a three dimensional network of $RhGe_4$ distorted tetrahedra, in which Rh-Ge bond distances are close to the covalent radii of the respective elements. The XRD pattern with Rietveld refinement for GdRhGe compound is shown in Fig. 1.

The magnetic susceptibility (χ) was measured in different fields in the temperature range of 1.8-300 K (see Fig. 2a). The χ vs. T plot obtained at 500 Oe shows a transition ($T_1$) at 31.8 K. On decreasing the temperature, there is broad peak around 24 K (termed as $T_2$), which becomes sharp on the application of 20 kOe field. It was observed that the peak temperature of first transition decreases slightly with increase in field, which suggests that the nature of first transition is antiferromagnetic. The Curie-Weiss law, $\chi^{-1} = (T - \theta_p)/C_m$ was fitted (see Fig. 2b) in the paramagnetic region. The effective magnetic moment ($\mu_{eff}$) and the paramagnetic Curie temperature ($\theta_p$), obtained from this fit are 8.5 $\mu_B/Gd^{3+}$ and -4.9 K. The value of observed $\mu_{eff}$ is higher than the theoretical value of 7.94 $\mu_B$, which may be either due the contribution of the moment on Rh and/or due to the strong polarization of conduction electrons by Gd [14]. The compound has negative $\theta_p$, which suggests that the antiferromagnetic ordering is predominant.

Since Gd is strong absorber of neutrons, it is not easy to carry out neutron diffraction experiment on this compound to find out the magnetic structure. However, some alternative measurements could give some clue about the magnetic phase transition in this compound. The ac susceptibility (ACS) is one such method, which gives information about the magnetic phase transition and the magnetization dynamics. The method has the advantage that it is done with small amplitudes of oscillating field, which does not perturb the magnetic system. Therefore, ACS measurement at different frequencies (f=25-450 Hz) for constant ac field ($H_{ac}$=1 Oe) in the temperature range of 1.8- 50 K has been performed for GdRhGe (see Fig. 3). Two well defined magnetic transitions in the real component ($\chi'$) of ACS coincide with the transition temperatures seen in the $\chi_{dc}$-T plot. However, one can see that there is only one peak in the imaginary part of ACS, corresponding to $T_2$ (shown by arrow). Furthermore the $\chi''$ increase significantly with decrease in temperature below $T_2$. Such maxima in $\chi''$ reflects energy losses in the magnetically ordered regime and arises due to the magnetic domain wall motion or the domain rotation in ferro/ferri magnets, spin glass or canted systems [15, 16]. On the other hand, a simple antiferromagnetic transition generally does not show up in the temperature dependence of $\chi''$ data [17]. Therefore, the ACS data in the present case suggest that the transition at $T_2$ is not a simple antiferromagnetic transition but it is a complex one while the transition at $T_1$ is simple antiferromagnetic (collinear).

Fig. 4 presents the magnetization isotherms at selected temperatures for fields up to 70 kOe. One can see that below a critical field $H_C$, the magnetization increases with increase in temperature up to $T_1$ and above $T_1$, it decreases with increase in temperature. Above the critical field, the trend is reversed, i.e., the magnetization decreases with increase in temperature. Hence it is clear that below $H_C$, the behavior is as expected for an antiferromagnet, which also shows a field-induced metamagnetic transition. The $H_C$ value required for the phase transition has been derived from the dM/dH plot as shown (for T=2 K) in the inset of Fig. 4. One can observe from this figure that the magnetization isotherms are almost linear; there is no indication of saturation of the moment. Further, no hysteresis has been detected in these isotherms. Hence, one can say that even after the metamagnetic transition, the magnetic state is far away from ferromagnetic.

In order to understand the system in more detail, the heat capacity measurements were performed with and without field in the temperature range of 2-100 K (see Fig. 5). One can see from heat capacity plot that there are λ- like peaks associated with both the transitions (at $T_1$ and $T_2$), implying that the magnetic ordering is of second order type. Both the peak heights get reduced on application of field. It is worth noting that the peak at $T_1$ shifts towards lower temperature, indicating that the transition is indeed antiferromagnetic, while there is no shift for the other peak, which is smeared out at 50 kOe.

To get a better insight into the magnetic state of the compound, electrical resistivity measurement was also carried out. Fig. 6a shows the variation of electrical resistivity with temperature for different fields. The linear behavior at high temperatures and the positive temperature coefficient of resistivity indicate the metallic nature of the compound. The slope change in resistivity has been observed at $T_1$ and $T_2$ (see inset in Fig. 6a), consistent with the magnetization and heat capacity data. To know the dominant contributions in the electrical resistivity data, the zero field resistivity data was fitted in low and high temperature regions using the well known theoretical models. Inset in Fig. 6b shows the low temperature fit using the equation [18],

$$\rho(T) = \rho_0 + AT^n \qquad (1)$$

Here, $\rho_0$ is residual resistivity. The value of $n$ evaluated from the fit is 2, which is generally observed for the ferromagnetic materials and suggests that spin wave scattering is dominant at low temperatures [18-20]. Therefore, it is quite evident from the fit that the magnetic phase below $T_2$ is not a simple antiferromagnetic one. The $\rho_0$ and A values are estimated to be $5.23 \times 10^{-6}$ ohm-cm and $8.51 \times 10^{-8}$ ohm-cm/$K^2$ for GdRhGe. The high temperature resistivity data was fitted using the equation

$$\rho(T) = B + CT - DT^3 \qquad (2)$$

Fig. 6b shows the fit of equation (2) in the high temperature regime. A good fit suggests that the resistivity at high temperatures is determined by the s-d scattering [20]. The estimated values of B, C and D are $2.58 \times 10^{-5}$ ohm-cm, $6.23 \times 10^{-7}$ ohm-cm/K, and $3.96 \times 10^{-13}$ ohm-cm/$K^3$ respectively.

The magnetoresistance has been calculated from the field dependence of the resistivity data using the formula, $MR\% = \left(\dfrac{\Delta\rho(T,H)}{\rho(T,0)}\right) \times 100$, where $\Delta\rho$ is the change in the zero field resistivity by the applied field. Fig. 7a shows the MR variation for a field change of 50 kOe. The value of MR in the paramagnetic regime is not significant. On lowering the temperature, near the transition temperature $T_1$, it shows a nominal negative MR. Below $T_1$ (i.e. at 31 K), the sign of MR changes and becomes positive. It is significant to note that, close to $T_2$, there is a visible change in the trend of MR, both in the temperature dependence and in the field dependence. On decreasing the temperature below $T_2$, the MR becomes strongly positive and attains a value of ≈48 % at 2 K. It can be seen from MR isotherms (see Fig. 7b) that, as the temperature is reduced below $T_1$, the sign of MR reverses and the magnitude increases linearly with field. It is important to note that large positive MR is observed over the metamagnetic region of the H-T space. It is of importance to mention here that, based on similar investigations that we have undertaken on other members of the RRhGe series, there is a general trend towards positive MR at low temperatures. It is true that the magnitudes of positive MR are different in different members of the series and also that in certain cases the positive MR is seen only in lower fields. In these latter cases, on application of 50 kOe field, the MR is found to become negative. Therefore, we report here that among the RRhGe series of compounds, GdRhGe presents the best scenario as far as large positive MR is concerned.

Generally, the occurrence of positive GMR can be explained by the Lorentz contribution to the excess resistivity. It has been found that single crystals of pure metals in which the low temperature resistivity is of the order of a few *nΩ-cm* generally show large positive MR [9]. But in our case the resistivity at low temperatures is of the order of a few μΩ-cm and therefore, the Lorentz contribution may not be the reason for the large positive MR value. Mazumdar *et al.* [9] have reported very large values of positive MR in antiferromagnetic $R_2Ni_3Si_5$ (R=Tb,Sm,Nd) compounds. This has been interpreted as a consequence of the layered structure containing sheets of rare earth ions [21]. However, in the present case, the structure is not a layered one. Very large positive MR was also observed in the antiferromagnetic $GdCu_6$ [11] in the ordered regime and nominal positive value in the paramagnetic regime. The authors have reported that the large positive MR in ordered regime might be due to either the enhancement of the spins in one of the magnetic sublattices or due to the reflection of *s* electrons from the potential barrier at the interphase boundary separating the phases with different magnetic structures. Unusually large positive MR at low temperatures was also reported by Nagasawa [12] in pure Nd metal, which possesses complex magnetic ordering with two antiferromagnetic transitions. Yamada and Takada [22] have attributed this to the inability of the applied field to suppress the spin disorder scattering effectively. Based on these observations and taking into account the M-T data, we feel that the scenario in GdRhGe is more or less similar to that of Nd. It may be recalled that, like in Nd metal, in GhRhGe also the exchange interaction is purely of RKKY type. This causes incommensurate/complex magnetic structure of this compound (below $T_1$). It may also be noted that the effect of field on the heat capacity peaks at $T_1$ and $T_2$ are different. Furthermore, the low temperature resistivity fit did not indicate an antiferromagnetic state. Therefore, it appears that the nature of magnetic order around $T_2$ is different from that just below $T_1$. This kind of a complex magnetic state would create a situation similar to that in Nd and cause an increase in resistivity with field. Another contribution that would aid the positive MR is the spin fluctuations arising from the 4d sublattice of Rh. Though no stable moment could be assigned to Rh from the magnetization data, the presence of fluctuating 4d moments of Rh cannot be ruled out. Application of a field would cause an enhancement of these fluctuations because of the large delocalization of the 4d band. Therefore, we feel that the sign reversal and the large value of MR observed in GdRhGe are closely related to the peculiar magnetic state of this compound and to a large extent true for the entire series.

In view of the large value and the sign reversal of MR seen in this compound, we have carried out the magneto-thermal study by probing the magnetocaloric effect. In the present case MCE has been calculated from M-H-T (M is magnetization and H is applied field) data. Usually, the MCE is expressed in terms of isothermal magnetic entropy change ($\Delta S_M$). In order to find out the magnetocaloric properties of GdRhGe, we have calculated $\Delta S_M$ using the Maxwell's relation, $\Delta S_M = \int_0^H \left[ \frac{\partial M}{\partial T} \right]_H dH$. Though the value of MCE calculated is not very high, it can give information about the nature of magnetic ordering, metamagnetic transitions etc [23]. It is clear from Fig. 8 that $\Delta S_M$ is negative (positive MCE) near $T_1$ and changes from negative to positive (negative MCE) when moving towards $T_2$. Hence, one can see that, like MR, $\Delta S_M$ also undergoes a sign reversal below $T_1$. Obviously, this must be related to the magnetic transition occurring at temperatures close to $T_2$. It may also be noticed that for a given field, the width of the peak at $T_2$ is more than that at $T_1$, reflecting the complex magnetic state around $T_2$.

## 4. Conclusions

In summary, we find that GdRhGe shows two magnetic transitions as revealed by the magnetization, the heat capacity and the resistivity data. Though both these transitions appear to be antiferromagnetic in nature, the low temperature transition is a result of some complex magnetic ordering as indicated by the magnetic susceptibility, magnetoresistance and magnetocaloric data. The sign reversal seen in the temperature variation of MR and MCE data also seems to be a consequence of the change in the magnetic structure below and above $T_1$. However, the contribution from a change in the amplitude of the moment with field in this temperature regime towards the sign reversal cannot be ruled out.

## Acknowledgments

S. G. thanks C.S.I.R., New Delhi for granting a fellowship. We acknowledge the help by D. Buddhikot in resistivity measurements.

**Figure Captions:**

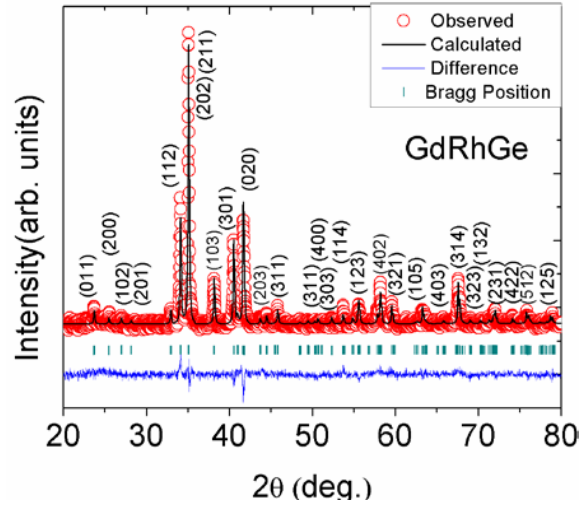

**Fig. 1.** Powder XRD pattern along with the Rietveld refinement for GdRhGe.

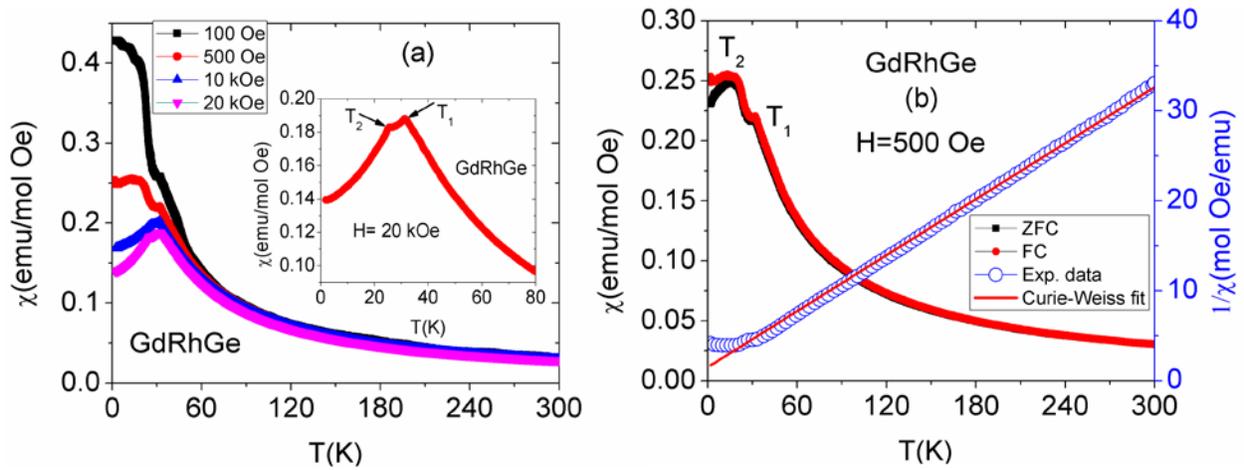

**Fig. 2.** (a) Temperature dependence of magnetic susceptibility obtained in various fields for GdRhGe. (b) Temperature dependence of the susceptibility obtained at 500 Oe. The right-hand scale is for the inverse susceptibility data. The inset in (a) shows an expanded plot for a field of 20 kOe.

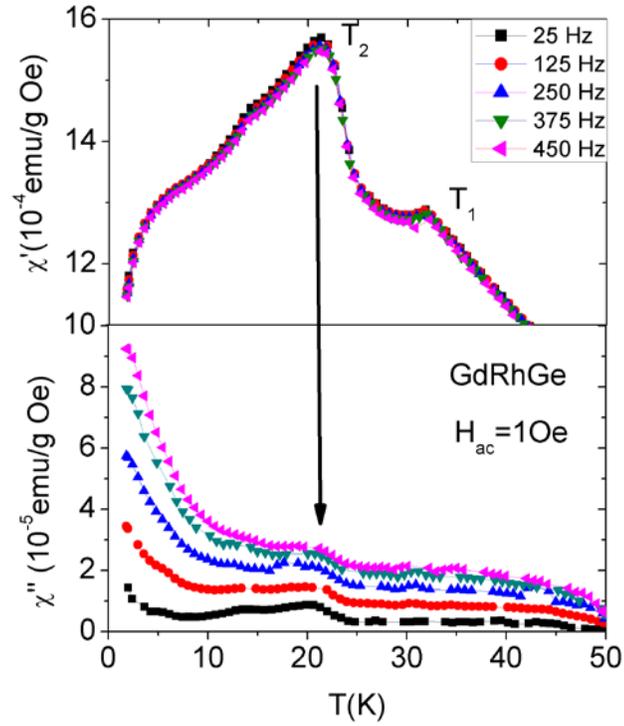

**Fig. 3.** Temperature dependence of ac magnetic susceptibility at various frequencies (f=25-450 Hz) and constant ac field ($H_{ac}$= 1Oe).

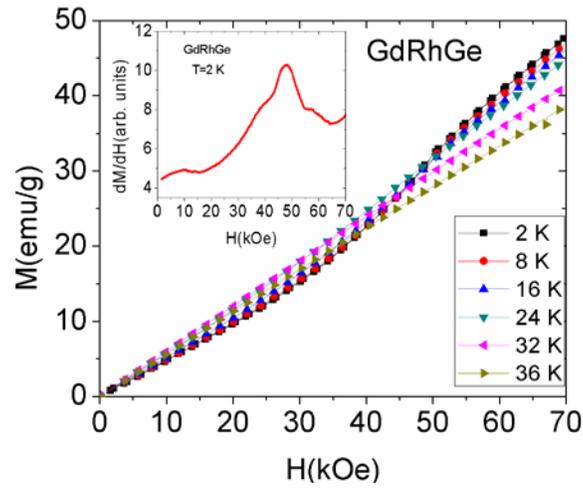

**Fig. 4.** Magnetization isotherms of GdRhGe at different temperatures. The inset shows the derivative of the magnetization with respect to the applied field.

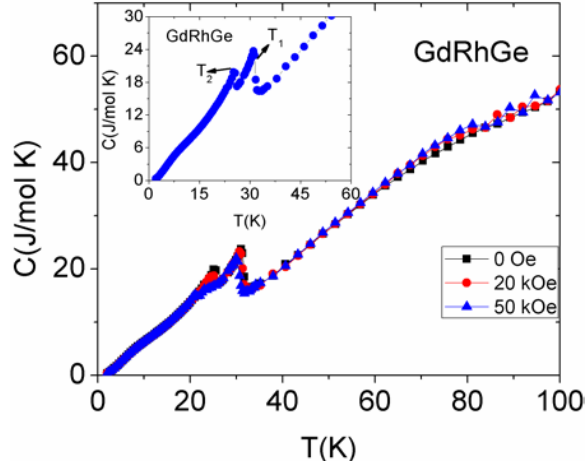

**Fig. 5.** C vs. T plots of GdRhGe in different fields. The inset shows an expanded plot of heat capacity at zero field.

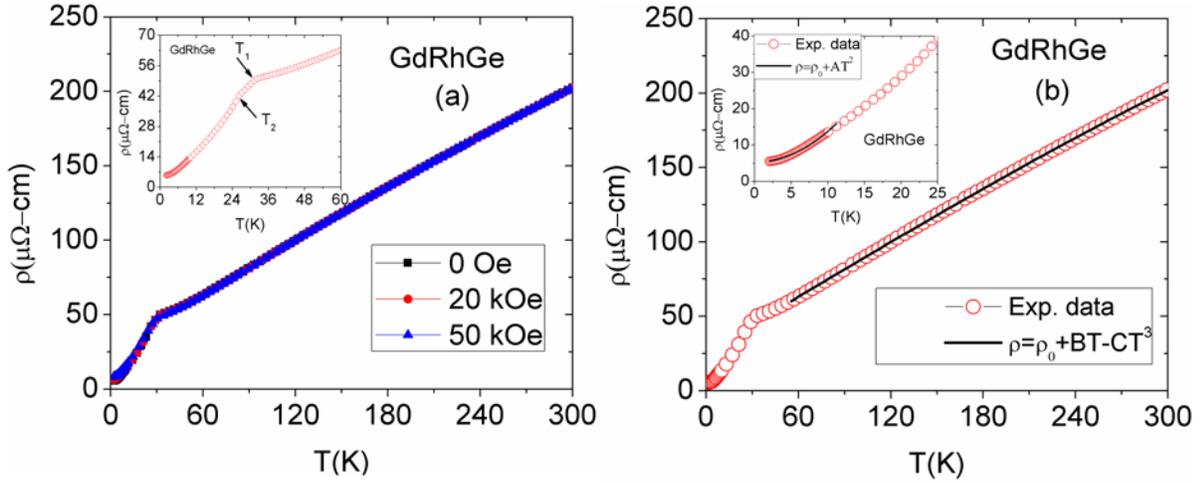

**Fig. 6.** (a) Temperature dependence of electrical resistivity at different fields in GdRhGe. The inset shows an expanded plot at low temperatures. (b) fit to the high temperature resistivity data. Inset in (b) shows the fit at low temperatures.

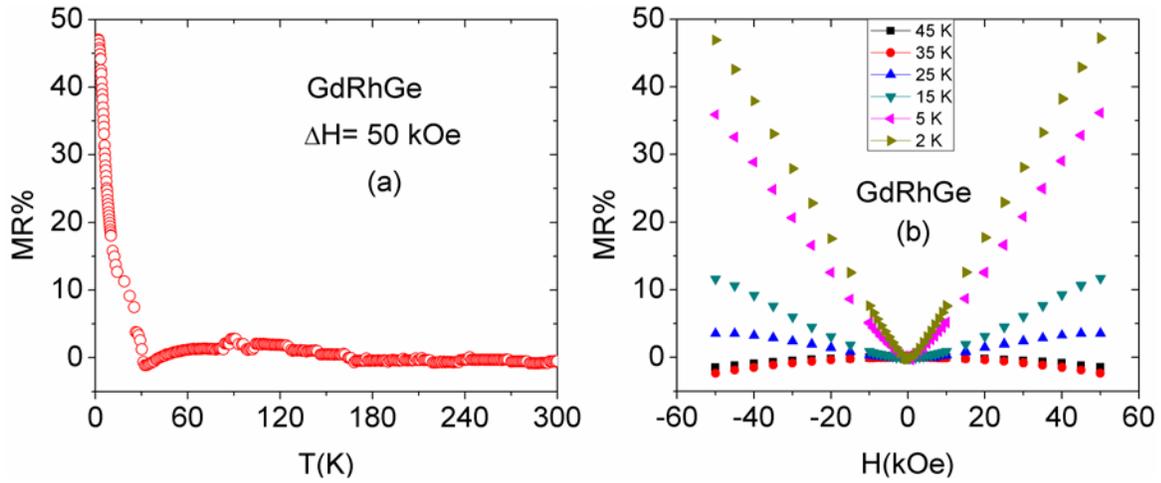

**Fig. 7.** (a) Temperature dependence of MR for a field change of 50 kOe, and (b) field dependence of MR at different temperatures.

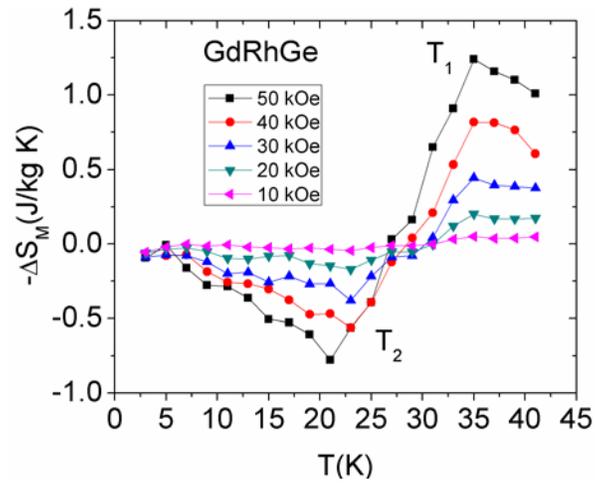

**Fig. 8.** Temperature variation of magnetic entropy change at various fields estimated from magnetization data.

13